# Electronic properties of carbon nanostructures based on bipartite nanocages units.

Fábio Nascimento de Sousa,[1,2] Divino Eliaquino,[3] Fabrício Morais de Vasconcelos,[4] and Eduardo Costa Girão[1,3]

[1]*Programa de Pós-Graduação em Ciência e Engenharia de Materiais, Universidade Federal do Piauí, CEP 64049-550, Teresina, Piauí, Brazil*

[2]*Instituto Federal de Educação, Ciência e Tecnologia do Piauí - Campus Teresina, CEP 64000-040, Teresina, Piauí, Brazil*

[3]*Departamento de Física, Universidade Federal do Piauí, CEP 64049-550, Teresina, Piauí, Brazil*

[4]*Instituto Federal de Educação, Ciência e Tecnologia do Piauí - Campus São João do PI, CEP 64760-000, São João do PI, Piauí, Brazil*

We use first principles simulations to investigate the electronic properties of a set of carbon nanocages with a bipartite structure. These nanocages are exclusively formed by hexagonal and tetragonal rings and we show they feature frontier states with particular symmetries as dictated by well defined rules based on the size of the structure. We also show that spin-polarized configuration emerge at the edges of open versions of these 0D systems. These cages are further proposed as elementary building blocks for periodic 1D and 2D systems. Even though we find different ways to define the particular bonds making the connection between the cage-like units, these systems always show a semiconducting behavior, both for 1D and 2D cases. However, the details of linking hierarchies interfere in the degree of localization of the frontier states of these crystalline systems.

## INTRODUCTION

The study of cage-like structures plays a relevant role in nanocarbon science. This is highlighted by the scientific and historical importance of the discovery of fullerenes [1], which can be viewed as the seminal mark in the study of nanoscaled carbon. Even though this mark is mostly associated to the C60 structure, such structure is only a member of a broad family of closed 0D nanocabons. Theoretical research further expanded to other cage structures with rings different from hexagons and pentagons [2–6]. In addition to cage geometries, closely related 0D carbon-cone and bowl-like geometries have also been considered, including theoretical studies [] and the experimental realization of some particular examples [7]. Examples include bowl systems similar to a fullerene hemisphere [8] and cones with angular aperture controlled by the number of pentagons at the cone tip [9]. In fact, carbon rings of order smaller than 6 are a basic ingredient to introduce positive curvature in these systems. Otherwise, higher-order rings, like pentagons induce negative curvature, producing saddle-shaped structures [10].

Recently, the C60 structure was involved in the construction of new 2D networks [11, 12]. The synthesis process involves the application of pressure and high temperatures on a mixture of $C_{60}$ and Mg powder to grow a magnesium-doped intermediate which was further processed to remove the dopants [12]. It has been shown that these structures can feature band gaps os about 1.6 eV [11]. Further theoretical research has explored chemical and physical properties of such kind of structures, including photocatalysis, mechanical and electronic properties [13–15]. This scenario has further motivated studies of 2D systems based on other cage-like structures [16]. In fact, the set of allowed geometries for 2D networks based on fullerenes and other cage structures is broad due mainly to two reasons. First, there are several different ways these units can be connected to each other. Second, the degree of freedom of choosing the geometry of the building block composing the system.

Here we study a set of systems based on a particular bipartite cage structure. Such cages are solely composed of tri-coordinated carbons forming either hexagons or squares and their size can be varied by the number of hexagons in the structure. A boron-nitride version of one of these systems has been previously studied [17]. However, to the best of our knowledge, there is no further study on full carbon versions of this structure. Initially we study the electronic properties of these cages for different sizes and a variation of these systems in form of nanocones. We show that different symmetries allowed for the frontier states wavefunctions dictate details of the electronic levels in the vicinity of the Fermi level. In the following, we propose two different strategies to concatenate these cages for form 1D tube-like and 2D membranes. In particular, we show that these systems are semiconducting and rationalize their properties in terms of the spatial distribution of their frontier states. The rest of this paper is as follows. In the next section we explain the structural details of these systems, including the 0D, 1D, and 2D forms. Then we dedicate a section for the employed methods. In the sequence we discuss our results and end the paper with our conclusions.

## STUDIED STRUCTURES

The number of rings in a cage-like structure solely composed by tri-coordinated carbon atoms can be determined with the aid of Euler's rule for a structure with genus zero ($g = 0$). With $N_i$ being the number of rings with $i$ sides, it is trivial to show that these quantities have to obey the following relation

$$2N_4 + N_5 - N_7 - 2N_8 - 3N_9 ... = 12(1-g). \quad (1)$$

In this work, we consider a particular 0D structural unit to construct different 1D and 2D systems. This corresponds to a cage-like structure solely composed by tricoordinated carbon atoms forming either tetragonal or hexagonal rings. In order to form a cage structure, it is necessary to have rings with less sides than a hexagon, since these polygons will introduce the positive curvature to close the cage structure. In a C60 fullerene, we have twelve pentagons, while here we consider tetragons. Using Eq. 1, we easily note that $N_4$ has to be 6, while $N_6$ is free. The resulting structures resemble octahedrons and one example is shown in Fig. 1a. In this example, there are two hexagons between two nearest tetragons, so that we call it a $n = 2$ structure. We will refer to these systems as *octahedron cages*, or simply OC-$n$. In the first part of this study, we consider cages with $n = 2$, ..., 7. The number of atoms ($N_{atoms}$) for a OC-$n$ is given by:

$$N_{atoms} = 8(n+1)^2. \quad (2)$$

We start with the $n = 2$ system because this is the system with the closest number of atoms (72) to a C60 fullerene. In the other extreme, we have $N_{atoms} = 512$ for $n = 7$. We further consider a variation of such structures consisting of bowl-like systems with hydrogen saturated edges, as illustrated in Fig. 1b for $n = 2$. This can be conceptually arranged by dividing the octahedron cage into two halves. We will refer to these systems as *half octahedron cages*, or simply HOC-$n$. It turns out these bowl systems will feature zigzag edges that are prone to show spin-polarized states, as observed for several graphitic systems with such kind of edge [18–21].

In the following, we consider quasi-1D nanotube-like structures composed of laterally fused cages. In the first case, we remove the atoms from two opposite squares of an OC-$n = 2$ to define the structural units. These are then concatenated successively by linking the dangling atoms from neighboring blocks. We call this system as a *octahedron carbon nanotube* of type $\alpha$ (or OCNT-$\alpha$). This nanotube structure is illustrated in Fig. 1c and it has a peapod-like form. In this way, the connection bewteen two successive OC-$n = 2$ units features negative curvature and we need polygons with more than 6 sides to close the lattice. In fact, besides hexagonal rings and four tetragons, the unit cell of this nanotube has a total of four octagons. Once the unit cell of a $sp^2$ nanotube system can be mapped into a torus structure (with unit genus - $g = 1$), the number of rings from each type in the system's unit cell has to obey

$$2N_4 + N_5 - N_7 - 2N_8 - 3N_9 ... = 0 \quad (3)$$

which is consisting with $N_4 = N_8 = 4$, as observed for the OCNT-$\alpha$. We consider a second nanotube-like structure (OCNT-$\beta$) in which no atoms are removed, but the tetragon atoms linking neighboring cage units are rearranged to form eight heptagons, which is consistent with the result $2N_4 = N_7$ from Eq. 3, since the tube still has a total of four tetragons in its unit cell. This nanotube structure is illustrated in Fig. 1d and it also features a peapod-like form, with negative curvature at the OC-$n = 2$ connections, hence the appearance of heptagons. In this part of the study, we only considered $n = 2$ blocks. In Fig. 1d,f we further illustrate amplified views of the links between two successive OC-2-like units, highlighting the octagons/heptagons in red color.

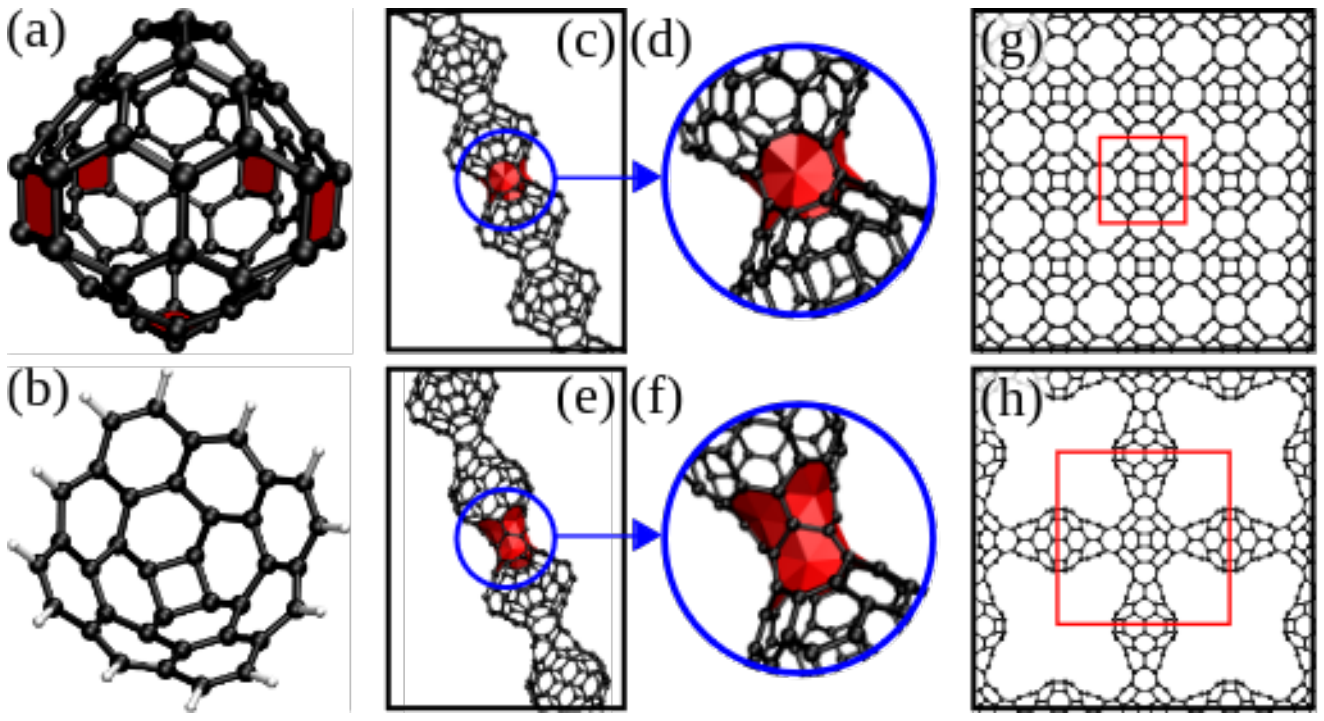

FIG. 1. (a) Representation of the atomic structure of a OC-2 structure, where the tetragonal rings are highlighted in red. Note that there is exactly 2 hexagons in between two neighboring squares. (b) Illustration of the HOC-2 variation of the OC-2 structure. (c) Illustration of the OCNT-$\alpha$ structure based on a OC-2-like building block. (d) Amplified vision of the junction between two successive OC-2-like units in the OCNT-$\alpha$. In (c-d), the octagons composing these junctions are highlighted in red color. (e-f) Same as (c-d), but for the OCNT-$\beta$ structure, where the heptagons composing the connection between successive OC-2-like units along the tube axis are highlighted in red color. (g) Atomic structure of the two-dimensional Sheet-$\alpha$ structure, where its tetragonal primitive unit cell is represented by the red square. (h) Same as (g), but for the Sheet-$\beta$ case.

Finally, we considered two quasi-2D systems consisting of $n = 2$ cage-like units linked together in square lattices by means of links similar to those involved in the definition of the OCNT-$\alpha$ and OCNT-$\beta$ structures. The corresponding Sheet-$\alpha$ and Sheet-$\beta$ structures are illustrated in Fig. 1g,h, respectively. Each of the Sheet-$\alpha$ and Sheet-$\beta$ structures contains a single hole per primitive unit cell. Since we can further map the unit cell of a 2D system into a torus topological counterpart, these systems can be described as $g = 2$ structures, so that

$$2N_4 + N_5 - N_7 - 2N_8 - 3N_9 ... = -12. \quad (4)$$

Besides featuring hexagons, Sheet-$\alpha$ has a total of 8 octagons and 2 tetragons per unit cell, consistent with $N_4 - N_8 = -6$, as obtained from Eq. 4. For Sheet-$\beta$, we has a total of 32 heptagons and 10 tetragons per unit cell, consistent with $2N_4 - N_7 = -12$ (also obtained from Eq. 4).

## METHODS

We proceeded with structural optimizations and electronic structure calculations by using density functional theory (DFT) [22, 23] as implemented in the SIESTA code [24]. We used a double-$\zeta$ localized basis set which includes an additional polarization function (DZP) in order to expand the wavefunctions of the valence functions. For the core electrons, we considered norm-conserving Troullier-Martins pseudopotentials [25]. The grid for real-space integrations was defined according to a 400 Ry cutoff energy, while integrations over the Brillouin zone were performed with a Monkhorst-Pack sampling. In order to have $k$-samplings with similar density of points for the different systems, we used 24 and 10 $k$-points for the OCNT-$\alpha$ and OCNT-$\beta$ systems, respectively. Conversely, the Sheet-$\alpha$ and Sheet-$\beta$ systems were described with $24 \times 24$ and $10 \times 10$ $k$-samples, respectively. Exchange-correlation energy was accounted for by means of the generalized gradient approximation (GGA) according to the Perdew-Burke-Ernzerhof parametrization [26]. Structural relaxations were performed with no constraints until the maximal force over any atoms become lower than 0.01 eV/Å, while the upper threshold tolerance for the stress components was set to 0.1 GPa.

## RESULTS AND DISCUSSION

### Nanocages

We start with the results for the OC-$n$ systems. In Fig. 2a we illustrate the atomic structures of the OCs with $n = 2$, ..., 7, together with a representation of the energy values for their electronic levels. In all these cases, the highest occupied molecular orbital (HOMO) is triply degenerated. The same occurs for the lowest unoccupied molecular orbital (LUMO) from each system. This is consistent with the symmetry of the OC structures, which is described by the $O_h$ point group. In fact, this group features four different irreducible representations with dimension 3, indicating that OC systems can have up to triply-degenerated states. The OC-2 has the widest HOMO-LUMO gap from the studied structures, 1.57 eV, with this value decreasing to 0.85 eV and 0.15 eV for the OC-3 OC-4 cases, respectively. When moving to OC-5, the gap increases to 0.91 eV, following a new decreasing for OC-6 and -7. This dependence of the HOMO-LUMO gap as a function of the size parameter $n$ is similar to the way the gap of armchair edged graphene nanoribbons (AGNRs) depend on the width, in terms of the number $N_{CC}$ of dimer lines along the ribbon's nonperiodic direction. Namely, these gaps can be assigned to three families of systems with $n = 3i + j$, where $i$ is an integer and $j = 0, 1, 2$ identifies the three families of behaviors.

We further extrapolated the relation between the HOMO-LUMO gap and the systems' size for a wider range of $n$ values by considering a first-nearest-neighbor tight-binding (TB) model. In Fig. 2 we show the HOMO-LUMO gap as a function of $n$ according to these TB simulations (black full lines with circles) for up to $n = 20$. We also superpose the DFT results for $n = 2$ to $n = 7$ (dashed green lines with circles). We note a qualitatively good agreement between TB and DFT for the smaller structures, as well as the "multiple of 3" behavior extends to larger systems.

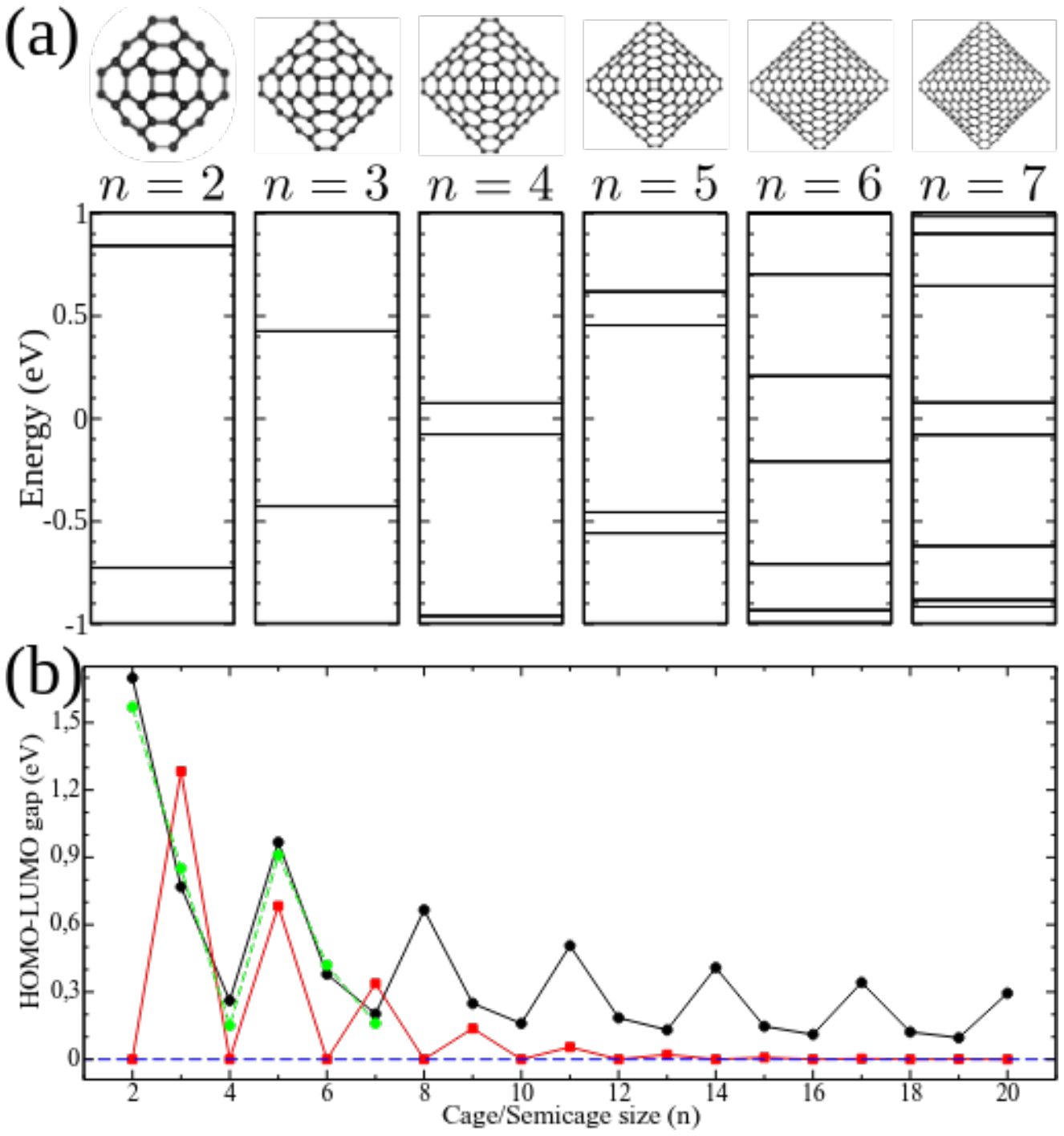

FIG. 2. (a) Illustration of the relaxed atomic structure of the OC-$n$ systems with $n = 2$, 3, 4, 5, 6, and 7, together with a representation of the values for their energy in the vicinity of the Fermi energy ($E = 0$). (b) Value of the HOMO-LUMO gap as a function of the system size parameter $n$ for the OC-$n$ systems with $n = 2$, 3, 4, 5, 6, and 7 according to DFT calculations (green line with circles), for OC-$n$ with $n = 2$, ..., 20 according to TBU calculations (black curve with circles), and for HOC-$n$ with $n = 2$, ..., 20 according to TBU calculations (red curve with squares).

In order to further investigate the "families of 3" behavior for the gap in these OC systems, we look at the symmetry of their frontier states. In order to simplify the analysis, we rely on the TB calculations, since the $\pi$-like states carry the major contribution for these levels. In terms of the $\pi$-like orbitals ($\phi_l$, where $l$ identifies the atomic site), the wavefunction $\psi_n$ corresponding to the energy level $E_n$ can be written as

$$\psi_n = \sum_l C_l^n \phi_l \qquad (5)$$

In Fig. 3 we represent the $C_l^n$ coefficients for the fron-tier states over the atomic structure in the OC-2, -3, -4, and -5 cases. In such representation, the size of the spheres are proportional to $|C_l^n|$, and positive (negative) coefficients are represented in blue (red) color. As discussed before, all these frontier states have triple degeneracy, compatible with the dimension of the irreducible representations of the $O_h$ point-symmetry group. In Table I we list the four tridimensional $O_h$ irreducible representations and corresponding sets of basis functions. In this table we also indicate when these basis functions are even/odd (+/-) or not symmetric (0) under selected symmetry operations of the $O_h$ group. These operations are illustrated in Fig. 3a, where $i$ represents the inversion operation, $x$ ($y$) [$z$] the plane orthogonal to the $Ox$ ($Oy$) [$Oz$] axis, $yz$ ($xz$) [$xy$] the plane bisecting the $y$ and $z$ ($x$ and $z$) [$x$ and $y$] planes, and $yz'$ ($xz'$) [$xy'$] the plane orthogonal to the $yz$ ($xz$) [$xy$] plane. These operations are enough to identify the symmetry of the frontier states in the OC systems. In Table II we list the HOMO and LUMO states of the OC-2, -3, -4, and -5 systems and the indication of when they are even/odd (+/-) or not symmetric (0) under the symmetry operations listed on Table I. By comparing Table I and Table II, we verify that moving from the HOMO to the LUMO in OC-2 corresponds to a transition from a $T_{2g}$ to a $T_{1u}$ state. The OC-3 has an extra hexagon between nearest tetragons in comparison to OC-2. This changes the way the positive and negative lobes of the HOMO and LUMO states are distributed over the structure. This includes changes in the position of the bonding and anti-bonding sectors of these states. These bonding and anti-bonding sections correspond to neighboring wavefunction lobes with same and opposite signs, respectively. As a result, the symmetry of the OC-3's frontier states change and correspond now to a transition from a $T_{1u}$ to a $T_{2g}$ state, as seem by comparing Table I and Table II. Further modifications in the wavefunctions symmetries are induced by the changes in the connectivity of the atomic structure when moving from OC-3 to OC-4. The HOMO-LUMO gap now corresponds to a transition from a $T_{1g}$ to $T_{2u}$ state. Finally, the HOMO and LUMO states of the OC-5 system recover the same symmetries from the OC-2, which is in the same multiple-of-3 family. So, the different trends for the HOMO-LUMO gap in the OC symmetries follows the symmetries allowed by the atomic connections in each family of systems.

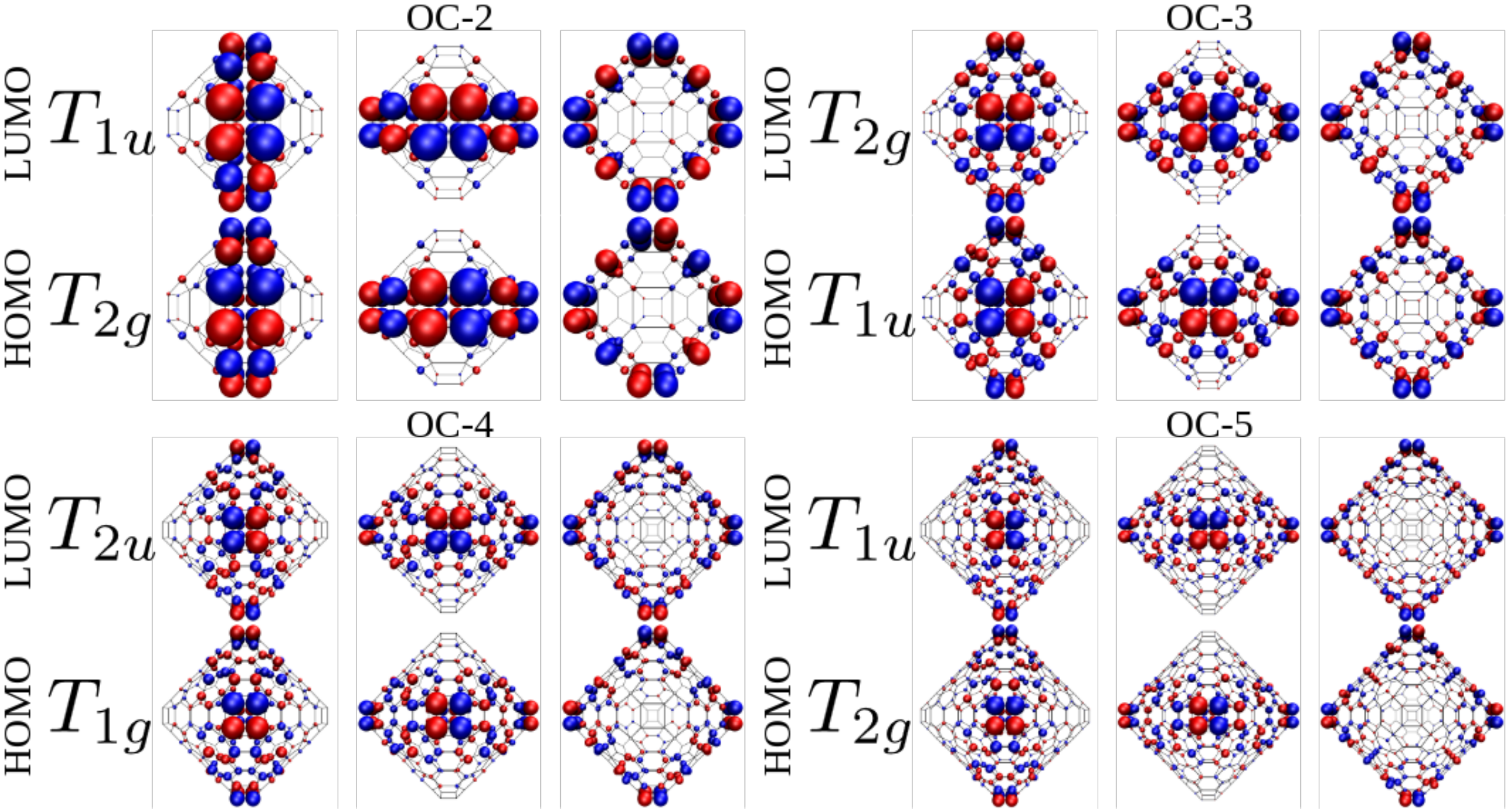

FIG. 3. Wavefunctions plots of the frontier HOMO and LUMO states for the OC-2, -3, -4, and -5 systems according to TBU simulations. In these representations, the diameter of the spheres is proportional to the absolute value of the coefficient of the $\pi$ orbital for each atomic site, while blue/red represent positive/negative values of the wavefunction coefficient. The symmetries of each set of states are also represented.

The electronic properties are significantly different if we cut the OC structures to form the bowl-like HOCs. In Fig. 4a we show the atomic structures of the SOCs with $n = 2$, ..., 7, together with a representation of the energy values for their electronic levels. As in the OC calculations, these results do not include the spin degree of freedom explicitly. Here, the HOMO-LUMO gap seems to oscillate between narrower (even $n$) and wider (odd

TABLE I. Tridimensional irreducible representations of the $O_h$ point-symmetry group, corresponding basis functions, and parity of each basis functions for selected symmetry operations from the $O_h$ group.

| $T_{1g}$ | $i$ | $x$ | $y$ | $z$ | $yz$ | $yz'$ | $xz$ | $xz'$ | $xy$ | $xy'$ |
|---|---|---|---|---|---|---|---|---|---|---|
| $yz(y^2-z^2)$ | + | + | - | - | - | - | 0 | 0 | 0 | 0 |
| $xz(x^2-z^2)$ | + | - | + | - | 0 | 0 | - | - | 0 | 0 |
| $xy(x^2-y^2)$ | + | - | - | + | 0 | 0 | 0 | 0 | - | - |

| $T_{2g}$ | $i$ | $x$ | $y$ | $z$ | $yz$ | $yz'$ | $xz$ | $xz'$ | $xy$ | $xy'$ |
|---|---|---|---|---|---|---|---|---|---|---|
| $yz$ | + | + | - | - | + | + | 0 | 0 | 0 | 0 |
| $xz$ | + | - | + | - | 0 | 0 | + | + | 0 | 0 |
| $xy$ | + | - | - | + | 0 | 0 | 0 | 0 | + | + |

| $T_{1u}$ | $i$ | $x$ | $y$ | $z$ | $yz$ | $yz'$ | $xz$ | $xz'$ | $xy$ | $xy'$ |
|---|---|---|---|---|---|---|---|---|---|---|
| $x$ | - | - | + | + | + | + | 0 | 0 | 0 | 0 |
| $y$ | - | + | - | + | 0 | 0 | + | + | 0 | 0 |
| $z$ | - | + | + | - | 0 | 0 | 0 | 0 | + | + |

| $T_{2u}$ | $i$ | $x$ | $y$ | $z$ | $yz$ | $yz'$ | $xz$ | $xz'$ | $xy$ | $xy'$ |
|---|---|---|---|---|---|---|---|---|---|---|
| $x(y^2-z^2)$ | - | - | + | + | - | - | 0 | 0 | 0 | 0 |
| $y(z^2-x^2)$ | - | + | - | + | 0 | 0 | - | - | 0 | 0 |
| $z(x^2-y^2)$ | - | + | + | - | 0 | 0 | 0 | 0 | - | - |

TABLE II. Parity of the HOMO and LUMO states of the OC-2, -3, -4, and -5 systems for selected symmetry operations from the $O_h$ group.

| OC-2 | $i$ | $x$ | $y$ | $z$ | $yz$ | $yz'$ | $xz$ | $xz'$ | $xy$ | $xy'$ |
|---|---|---|---|---|---|---|---|---|---|---|
| HOMO 3 | + | + | - | - | + | + | 0 | 0 | 0 | 0 |
| HOMO 2 | + | - | + | - | 0 | 0 | + | + | 0 | 0 |
| HOMO 1 | + | - | - | + | 0 | 0 | 0 | 0 | + | + |
| LUMO 1 | - | - | + | + | + | + | 0 | 0 | 0 | 0 |
| LUMO 2 | - | + | - | + | 0 | 0 | + | + | 0 | 0 |
| LUMO 3 | - | + | + | - | 0 | 0 | 0 | 0 | + | + |

| OC-3 | $i$ | $x$ | $y$ | $z$ | $yz$ | $yz'$ | $xz$ | $xz'$ | $xy$ | $xy'$ |
|---|---|---|---|---|---|---|---|---|---|---|
| HOMO 1 | - | - | + | + | + | + | 0 | 0 | 0 | 0 |
| HOMO 2 | - | + | - | + | 0 | 0 | + | + | 0 | 0 |
| HOMO 3 | - | + | + | - | 0 | 0 | 0 | 0 | + | + |
| LUMO 1 | + | + | - | - | + | + | 0 | 0 | 0 | 0 |
| LUMO 2 | + | - | + | - | 0 | 0 | + | + | 0 | 0 |
| LUMO 3 | + | - | - | + | 0 | 0 | 0 | 0 | + | + |

| OC-4 | $i$ | $x$ | $y$ | $z$ | $yz$ | $yz'$ | $xz$ | $xz'$ | $xy$ | $xy'$ |
|---|---|---|---|---|---|---|---|---|---|---|
| HOMO 1 | + | + | - | - | - | - | 0 | 0 | 0 | 0 |
| HOMO 2 | + | - | + | - | 0 | 0 | - | - | 0 | 0 |
| HOMO 3 | + | - | - | + | 0 | 0 | 0 | 0 | - | - |
| LUMO 1 | - | - | + | + | - | - | 0 | 0 | 0 | 0 |
| LUMO 2 | - | + | - | + | 0 | 0 | - | - | 0 | 0 |
| LUMO 3 | - | + | + | - | 0 | 0 | 0 | 0 | - | - |

| OC-5 | $i$ | $x$ | $y$ | $z$ | $yz$ | $yz'$ | $xz$ | $xz'$ | $xy$ | $xy'$ |
|---|---|---|---|---|---|---|---|---|---|---|
| HOMO 1 | + | + | - | - | + | + | 0 | 0 | 0 | 0 |
| HOMO 2 | + | - | + | - | 0 | 0 | + | + | 0 | 0 |
| HOMO 3 | + | - | - | + | 0 | 0 | 0 | 0 | + | + |
| LUMO 1 | - | - | + | + | + | + | 0 | 0 | 0 | 0 |
| LUMO 2 | - | + | - | + | 0 | 0 | + | + | 0 | 0 |
| LUMO 3 | - | + | + | - | 0 | 0 | 0 | 0 | + | + |

$n$) values. The $n = 6$ structure, in particular, shows a pair of $E = E_F$ states (with $E_F$ denoting the Fermi level). This trend is further verified for broader $n$ range by TB calculations, as illustrated by the red line with squares in Fig. 2. This change in comparison to the OCs case can be understood in terms of group theory. The symmetry of the HOC structures is now described by a less symmetric group, $C_{4v}$. This groups has only 1D and 2D irreducible representations. We note that the frontier states of the HOC-$n$ systems with odd $n$ are always doubly-degenerated, so that they are necessarily $E$ states (which is the only 2D irreducible representation of $C_{4v}$). On the other hand, the symmetry of the frontier HOMO and LUMO states in even $n$ OCs changes due to the increasing of hexagonal rings in the system and become nondegenerated levels.

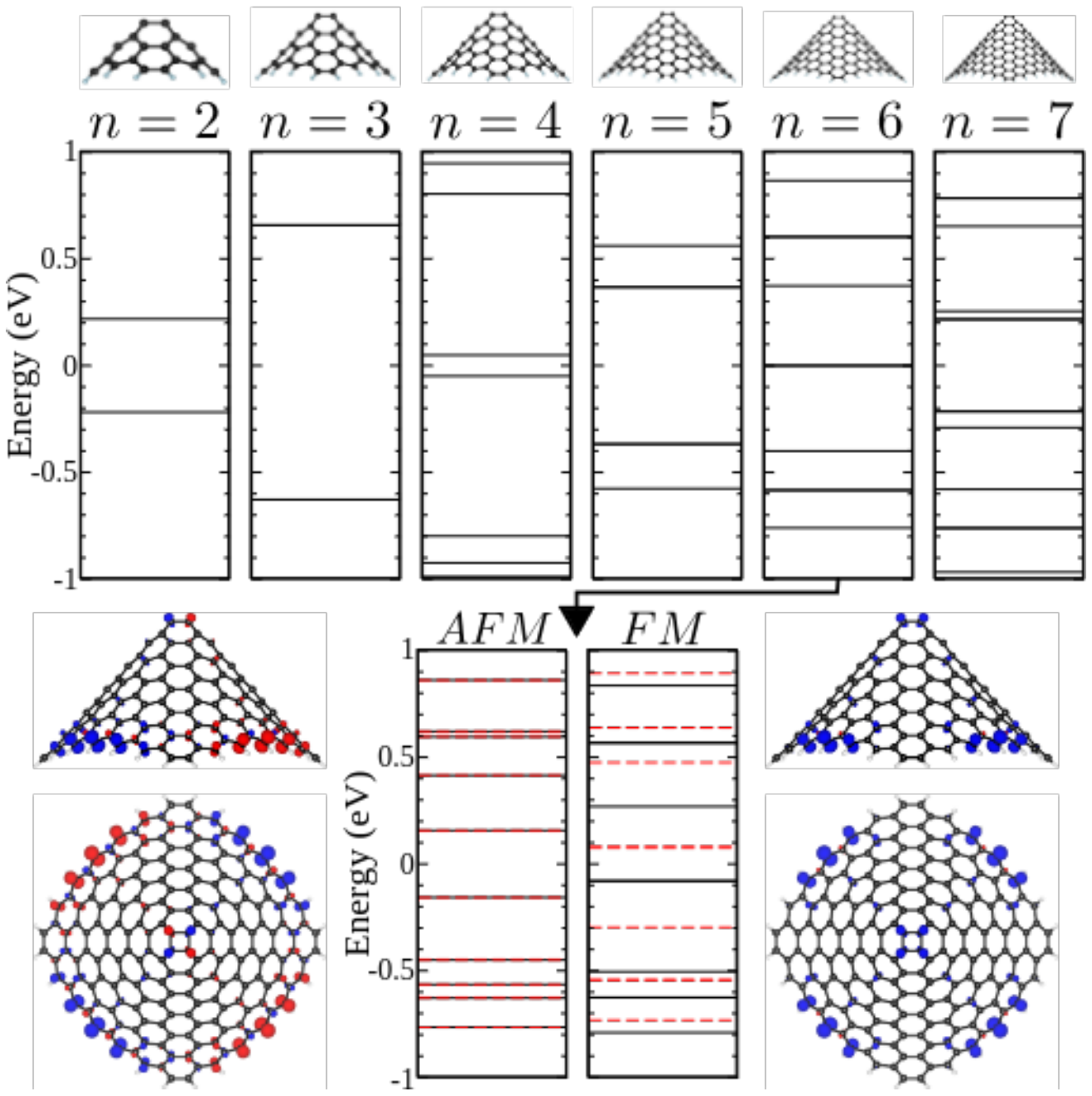

FIG. 4. (a) Illustration of the relaxed atomic structure of the HOC-$n$ systems with $n$ = 2, 3, 4, 5, 6, and 7, together with a representation of the values for their energy in the vicinity of the Fermi energy ($E = 0$). (b) Spin polarized AFM and FM states of the OC-6 system, where spin-up (-down) levels are represented by blue full (red dashed) lines. We also represent the difference between the spin-up and -down components of the total charge over this structure in the AFM and FM cases both in a side and upper view.

On the other hand, zigzag edged structures usually host spin-polarized states. Such a trend can be inferred when the electronic configuration with compensated spins shows low energy levels as those from HOC-6. We further performed spin-polarized calculations for this system. We observe a pair of spin-polarized configurations. They differ mostly by the relative alignment of the spins between neighboring edges. In the first case, the four system's edges alternate between majority spin-up and -down components. We call this configuration as anti-ferromagnetic (AFM). In the second case, all the four edges have the same majority spin, so that we call it a ferromagnetic (FM) configuration. Both AFM and FM open a HOMO-LUMO gap compared to the spin-paired configuration, which has zero gap. The AFM gap is xx eV, while the FM case shows a narrower gap which is between a spin-up and a spin-down state.

### 1D and 2D structures

We now move to the results for the OCNT structures. The lattice parameters for the $\alpha$ and $\beta$ cases are 8.03 Å and 19.90 Å, respectively. The $\beta$ cell is naturally longer since no atoms are removed from the structure at the squares positions. We define the width for these OCNTs by the distance between the centers of two squares on opposite sides of the tube along a direction orthogonal to its axis. These are very similar to the corresponding distance in OC-2, namely 8.34 Å for OCNT-$\alpha$, -$\beta$, OC-2. In Fig. 5 we show two different side views os each of the $\alpha$ and $\beta$ systems, together with the corresponding band structures.

The $\alpha$ system has an indirect gap of 0.93 eV (from $\Gamma$ to $X$), even though the direct gap at the $X$ point is only 36 meV wider. This difference corresponds to the bandwidth of the valence band, which is particularly low dispersive. We further plot the local density of states (LDOS) for the valence band maximum (VBM) and conduction band minimum (CBM) of this tube, as shown in Fig. 5. These two states are segregated from each other and are not strongly delocalized along the length of the system. The VBM is particularly distributed over the squares of an OC-2 unit and the hexagons connecting them, while they show negligible amplitude over the connection between successive OC-2 blocks (the octagonal rings). The CBM is mostly spread over the regions where the VBM does not show significant amplitude, namely over the octagons between two consecutive OC-2-lie units. However, the CBM features larger delocalization than the VBM, as it spreads significantly outside the octagons. This is consistent with the conduction band profile, which is much more dispersive than the valence band.

On the other hand, OCNT-$\beta$ features a direct gap of 1.11 eV at the $\Gamma$ point between two bands of similar dispersion. These states are also doubly degenerated at the edge of the Brillouin zone, with a folding-like pattern. This is associated with the screw axis symmetry of the structure, as verified for other 1D systems with this kind of symmetry. We also plot the LDOS for the VBM and CBM states in Fig. 5. These corresponding clouds have a delocalized character, differently than the VBM states of the OCNT-$\alpha$.

It is further interesting to compare the gaps for these tubes with those of conventional carbon nanotubes based on the graphene lattice. In particular, the (6,6) and (11,0) nanotubes have diameter similar to the width of the two studied OCNTs, namely 8.29 Å and 8.68 Å, respectively. While the (6,6) tube is metallic, the (11,0) case is semiconducting with a band gap of 0.88 eV.

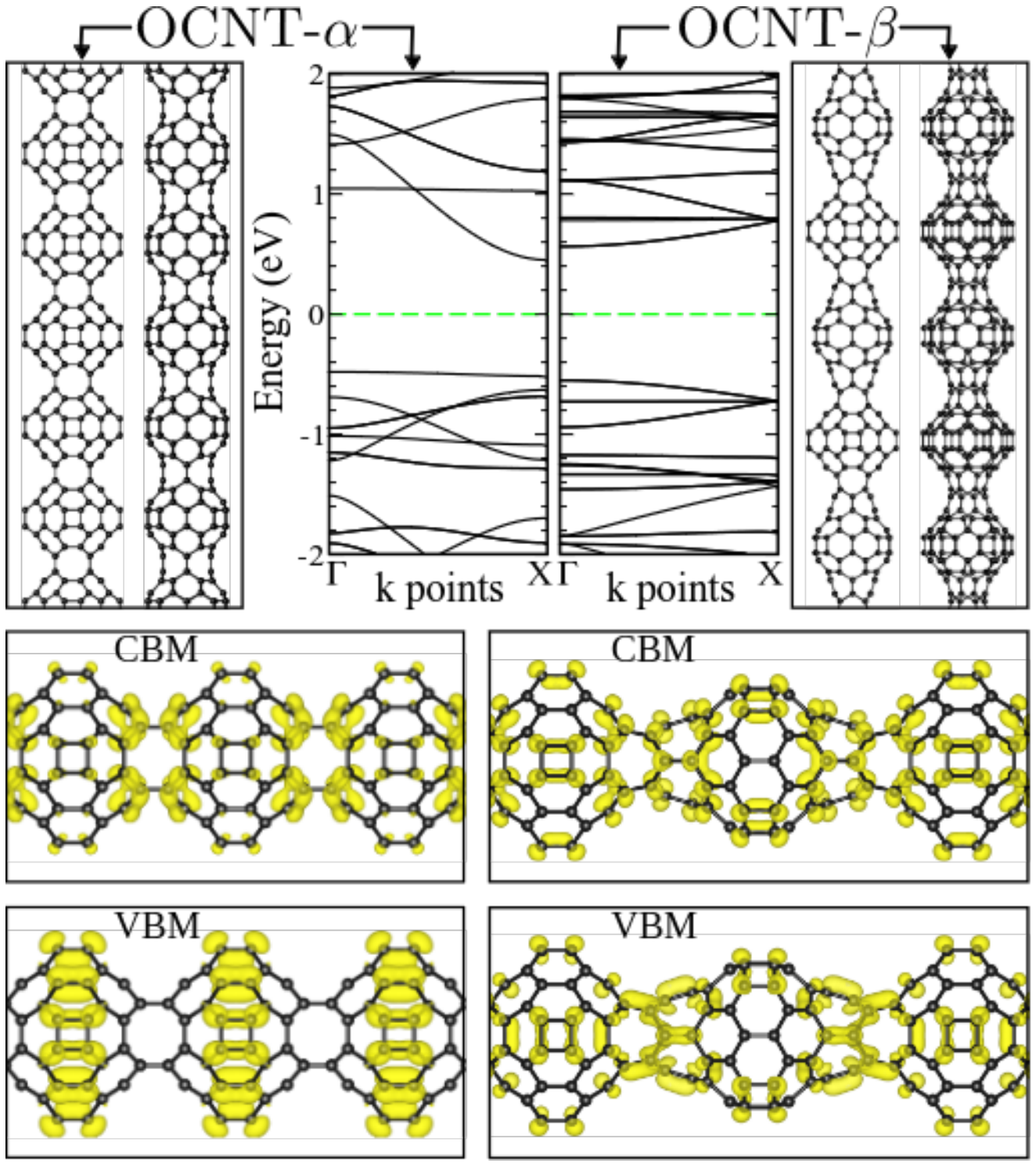

FIG. 5. (Upper panel) Two different side views and the electronic band structure of the OCNT-$\alpha$ and OCNT-$\beta$ systems. (Lower panel) LDOS plots for the VBM and CBM states of these two systems.

In the following, we show the electronic structure results for the 2D systems. The electronic band structure of Sheet-$\alpha$ is shown along the high-symmetry lines of its quadrangular Brillouin zone in Fig. 6. This is a semiconducting system with a 1.28 eV gap, about 38 % wider than the corresponding nanotube structure. One could expect the gap of the 2D system to be wider because it is a more uniform system than a nanotube. The 2D structure would favor delocalization, resulting in larger dispersion and narrower gaps. However this is the opposite to what we observe for Sheet-$\alpha$. We can understand this by looking at the flat valence band of OCNT-$\alpha$ and at the occupied bands of the 2D system. Onde the structure is cast into the 2D form, the corresponding flat level is shifted down, as we can see a low dispersive band along the $\Gamma - X$ path of the BZ around -1.06 eV. As a result, the more dispersive band below the flat valence band of the nanotube is promoted to valence band in Sheet-$\alpha$. In Fig. 6 we further plot the LDOS for the VBM and CBM of Sheet-$\alpha$, where we note that the VBM is delocalized over the structure (consistent with a dispersive band), differently from the VBM in OCNT-$\alpha$.

Moving to the Sheet-$\beta$ case, its gap is 0.94 eV, nearly 15 % narrower than the gap of the corresponding OCNT-$\beta$. The electronic band structure of Sheet-$\beta$ is shown in Fig. 6, together with the LDOS plots for the corresponding VBM and CBM. We note the LDOS clouds for the frontier states resemble those from the corresponding states in OCNT-$\beta$. In this way, the gap variation from OCNT-$\beta$ to Sheet-$\beta$ is mostly due to the rehybridization of the frontier states due to the larger number of OC-2 connections so that the gaps of these two systems are closer to each compared to the $\alpha$ systems.

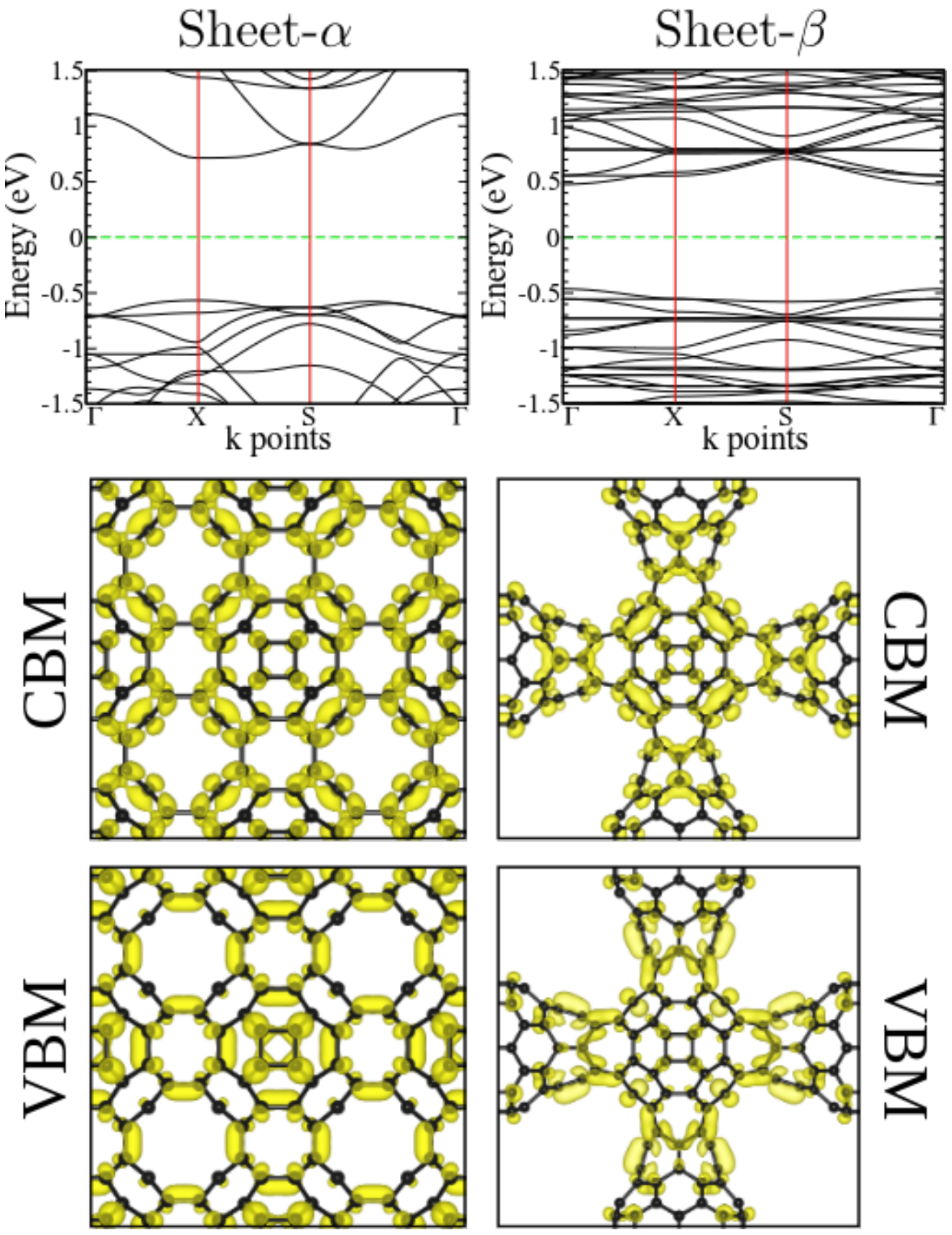

FIG. 6. (Upper panel) Electronic band structure of the Sheet-$\alpha$ and Sheet-$\beta$ systems along the high-symmetry lines of the Brillouin zone. (Lower panel) LDOS plots for the VBM and CBM states of these two systems.

## CONCLUDING REMARKS

In summary, we studied a series of cage-like nanocarbons with a bipartite structure. The 0D systems feature a HOMO-LUMO gap which is determined by the system's size according to a hierarchical "multiple-of-3" rule dictated by the accommodation of different symmetries for the frontier states. We further used the smallest studied cage as a basic building block to assembly 1D and 2D systems by means of different linking details. These periodic systems are all semiconducting and the degree of spatial delocalization of their frontier states is directly influenced by the details of the linking hierarchy.

## ACKNOWLEDGEMENTS

The authors thank the Laboratório de Simulação Computacional Cajuína (LSCC) at Universidade Federal do Piauí for computational support.

[1] H. W. Kroto, J. R. Heath, S. C. Obrien, R. F. Curl, R. E. Smalley. "C-60 - Buckminsterfullerene". *Nature* **318**(6042), 162 (1985).
[2] H. Terrones, M. Terrones. "Quasiperiodic icosahedral graphite sheets and high-genus fullereneswith nonpositive gaussian curvature". *Physical Review B* **55**, 9969 (1997).
[3] J. L. Ricardo-Chávez, J. Dorantes-Dávila, M. Terrones, H. Terrones. "Electronic properties of fullerenes with nonpositive gaussian curvature: Finite zeolites". *Physical Review B* **56**, 12143 (1997).
[4] H. Terrones, M. Terrones. "Fullerenes and nanotubes with non-positive gaussian curvature". *Carbon* **36**(5), 725 (1998).
[5] M. I. Heggie, M. Terrones, B. R. Eggen, G. Jungnickel, R. Jones, C. D. Latham, P. R. Briddon, H. Terrones. "Quantitative density-functional study of nested fullerenes". *Physical Review B* **57**, 13339 (1998).
[6] P. Schwerdtfeger, L. N. Wirz, J. Avery. "The topology of fullerenes". *WIREs Computational Molecular Science* **5**(1), 96 (2015).
[7] Q. Zhang, X.-M. Xie, S.-Y. Wei, Z.-Z. Zhu, L.-S. Zheng, S.-Y. Xie. "The synthesis of conical carbon". *Small Methods* **5**(3), 2001086 (2021).
[8] L. T. Scott, E. A. Jackson, Q. Zhang, B. D. Steinberg, M. Bancu, B. Li. "A short, rigid, structurally pure carbon nanotube by stepwise chemical synthesis". *Journal of the American Chemical Society* **134**(1), 107 (2012).
[9] Z.-Z. Zhu, Z.-C. Chen, Y.-R. Yao, C.-H. Cui, S.-H. Li, X.-J. Zhao, Q. Zhang, H.-R. Tian, P.-Y. Xu, F.-F. Xie, X.-M. Xie, Y.-Z. Tan, S.-L. Deng, J. M. Quimby, L. T. Scott, S.-Y. Xie, R.-B. Huang, L.-S. Zheng. "Rational synthesis of an atomically precise carboncone under mild conditions". *Science Advances* **5**(8), 982 (2019).
[10] J. M. Fernández-García, P. J. Evans, S. Medina Rivero, I. Fernández, D. García-Fresnadillo, J. Perles, J. Casado, N. Martín. "$\pi$-extended corannulene-based nanographenes: Selective formation of negative curvature". *Journal of the American Chemical Society* **140**(49), 17188 (2018).
[11] L. Hou, X. Cui, B. Guan, S. Wang, R. Li, Y. Liu, D. Zhu, J. Zheng. "Synthesis of a monolayer fullerene network". *Nature* **606**(7914), 507 (2022).
[12] E. Meirzadeh, A. M. Evans, M. Rezaee, M. Milich, C. J. Dionne, T. P. Darlington, S. T. Bao, A. K. Bartholomew, T. Handa, D. J. Rizzo, R. A. Wiscons, M. Reza, A. Zangiabadi, N. Fardian-Melamed, A. C. Crowther, P. J. Schuck, D. N. Basov, X. Zhu, A. Giri, P. E. Hopkins,

P. Kim, M. L. Steigerwald, J. Yang, C. Nuckolls, X. Roy. "A few-layer covalent network of fullerenes". *Nature* **613**(7942), 71 (2023).

[13] R. M. Tromer, L. A. Ribeiro, D. S. Galvão. "A dft study of the electronic, optical, and mechanical properties of a recently synthesized monolayer fullerene network". *Chemical Physics Letters* **804**, 139925 (2022).

[14] L. Ribeiro, M. Pereira, W. Giozza, R. Tromer, D. S. Galvão. "Thermal stability and fracture patterns of a recently synthesized monolayer fullerene network: A reactive molecular dynamics study". *Chemical Physics Letters* **807**, 140075 (2022).

[15] B. Peng. "Monolayer fullerene networks as photocatalysts for overall water splitting". *Journal of the American Chemical Society* **144**(43), 19921 (2022).

[16] B. Mortazavi, F. Shojaei, X. Zhuang. "A novel two-dimensional c36 fullerene network; an isotropic, auxetic semiconductor with low thermal conductivity and remarkable stiffness". *Materials Today Nano* **21**, 100280 (2023).

[17] M. D. Esrafili, R. Nurazar. "A density functional theory study on the adsorption and decomposition of methanol on b12n12 fullerene-like nanocage". *Superlattices and Microstructures* **67**, 54 (2014).

[18] Y.-W. Son, M. L. Cohen, S. G. Louie. "Half-metallic graphene nanoribbons". *Nature* **444**(7117), 347 (2006).

[19] L. Pisani, J. A. Chan, B. Montanari, N. M. Harrison. "Electronic structure and magnetic properties of graphitic ribbons". *Physical Review B* **75**(6), 064418 (2007).

[20] J. Fernández-Rossier, J. J. Palacios. "Magnetism in graphene nanoislands". *Physical Review Letters* **99**, 177204 (2007).

[21] O. V. Yazyev. "Emergence of magnetism in graphene materials and nanostructures". *Reports on Progress in Physics* **73**(5), 056501 (2010).

[22] P. Hohenberg, W. Kohn. "Inhomogeneous Electron Gas". *Physical Review B* **136**(3B), B864 (1964).

[23] W. Kohn, L. J. Sham. "Self-Consistent Equations Including Exchange and Correlation Effects". *Physical Review* **140**(4A), 1133 (1965).

[24] J. Soler, E. Artacho, J. Gale, A. Garcia, J. Junquera, P. Ordejon, D. Sanchez-Portal. "The siesta method for ab initio order-n materials simulation". *Journal of Physics-Condensed Matter* **14**(11), 2745 (2002).

[25] N. Troullier, J. L. Martins. "Efficient pseudopotentials for plane-wave calculations". *Physical Review B* **43**, 1993 (1991).

[26] J. P. Perdew, K. Burke, M. Ernzerhof. "Generalized gradient approximation made simple". *Physical Review Letters* **77**, 3865 (1996).